\documentclass[%
 aip,
 rsi,%
 amsmath,amssymb,
reprint,% paper
floatfix % ``emergency'' float placement
]{revtex4-1}

\usepackage{graphicx}% Include figure files
\usepackage{color}
\usepackage{dcolumn}% Align table columns on decimal point
\usepackage{bm}% bold math
\usepackage{placeins}

\begin{document}

\preprint{AIP/123-QED}

\title[Time-resolved measurement of the tunnel magneto-Seebeck effect in a 
single magnetic tunnel junction]{Time-resolved measurement of the tunnel 
magneto-Seebeck effect in a single magnetic tunnel junction}% Force line breaks with \\

\author{Alexander Boehnke}
\email{boehnke@physik.uni-bielefeld.de}
\affiliation{Thin Films and Physics of Nanostructures, Universit\"at Bielefeld,  
Universit\"atsstrasse 25, D-33615 Bielefeld,~Germany}
\author{Marvin~Walter}
\email{mwalter1@gwdg.de}
\author{Niklas Roschewsky}
\author{Tim Eggebrecht}
\affiliation{I. Physikalisches Institut, Georg-August-Universit\"at G\"ottingen, 
Friedrich-Hund-Platz 1, D-37077 G\"ottingen, Germany}%
\author{Volker Drewello}
\author{Karsten~Rott}
\affiliation{Thin Films and Physics of Nanostructures, Universit\"at Bielefeld,  
Universit\"atsstrasse 25, D-33615 Bielefeld,~Germany}
\author{Markus~M\"unzenberg}
\affiliation{I. Physikalisches Institut, Georg-August-Universit\"at G\"ottingen, 
Friedrich-Hund-Platz 1, D-37077 G\"ottingen, Germany}%
\author{Andy~Thomas}
\author{G\"unter Reiss}
\affiliation{Thin Films and Physics of Nanostructures, Universit\"at Bielefeld,  
Universit\"atsstrasse 25, D-33615 Bielefeld,~Germany}

\date{\today}% It is always \today, today,

\begin{abstract}
Recently, several groups have reported spin-dependent thermoelectric effects in magnetic 
tunnel junctions. In this paper, we present a setup for time-resolved measurements of 
thermovoltages and thermocurrents of a single micro- to nanometer-scaled tunnel junction.
An electrically modulated diode laser is used to create a temperature gradient across 
the tunnel junction layer stack. This laser modulation technique enables the recording 
of time-dependent thermovoltage signals with a temporal resolution only limited by the 
preamplifier for the thermovoltage. So far, time-dependent thermovoltage could not be 
interpreted. Now, with the setup presented in this paper, it is possible to distinguish 
different Seebeck voltage contributions to the overall measured voltage signal in the 
$\mathrm{\mu s}$ time regime. A model circuit is developed that explains those voltage 
contributions on different sample types. Further, it will be shown that a voltage signal 
arising from the magnetic tunnel junction can only be observed when the laser spot is 
directly centered on top of the magnetic tunnel junction, which allows a lateral 
separation of the effects.
\end{abstract}

			%Use showkeys class option if keyword
\maketitle

\section{Introduction}
\label{sec:introduction}
In recent years, the research field ``spin caloritronics'' has attracted considerable 
attention in the magnetism and spintronics communities\cite{Bauer2010,Bauer2012}. New 
spin-dependent thermoelectric effects have been discovered in ferromagnetic 
metals\cite{Uchida2008}, insulators\cite{Uchida2010} and 
semiconductors\cite{Jaworski2010}. Triggered by the experiments of Gravier et 
al.\cite{Gravier2006} and Shi et al.\cite{Shi1993} on giant magneto resistance 
(GMR) multilayers, and by theoretical predictions of large magnetothermoelectric effects 
in magnetic tunnel junctions\cite{Czerner2011}, several groups reported observations of 
a tunnel magneto-Seebeck effect (TMS) in magnetic tunnel junctions (MTJs) with 
MgO-\cite{Liebing2011,Walter2011} and alumina-barriers\cite{Lin2012}. A closely 
related effect is thermal spin injection into silicon through Seebeck spin 
tunneling\cite{LeBreton2011}. In non-local spin valves, thermally driven spin 
injection was discovered\cite{Slachter2010} and Peltier and Seebeck effects were 
studied\cite{Bakker2010}. The number of these new effects, combined with the proposed 
thermal spin-transfer torque \cite{Jia2011,Leutenantsmeyer2013} might enable the 
fabrication of thermally driven Magnetoresistive Random Access Memory (MRAM) and other 
spintronic devices.

Some of the effects are vividly discussed in the 
community.\cite{Kikkawa2013,geprags:262407,Huang2012,Meier2013,Qu2013,Avery2012,Lu2013}
In contrast, the experiments on CoFeB/MgO-based MTJs with high tunnel magnetoresistance 
(TMR) ratios, in which either a laser\cite{Walter2011} or resistive 
heating\cite{Liebing2011} is used to generate the temperature gradients, show comparable 
results. These are of the same magnitude as predicted by \textit{ab initio} 
calculations.\cite{Czerner2011,Walter2011} For MTJs with alumina-barrier, larger Seebeck 
voltages as compared to MgO-barriers at comparable temperature gradients are found. 
However, there are a few variations between the different experiments: First, the sign 
of the Seebeck voltage remains unclear, which could also vary depending on temperature 
and Co-Fe-composition.\cite{Heiliger2013}
Further, it was reported by Ref.~\onlinecite{Lin2012} that no magnetic effect is
observed in the thermocurrent obtained with alumina-barriers and that Seebeck voltages
could be observed when heating the electrical leads a distance of the order of
millimeters away from the MTJ.

In the following, we will address these issues and show results for the Seebeck voltage 
as well as for the thermocurrent and the determination of the voltage sign with a 
lock-in technique in section \ref{sec:experiments}. A model circuit is developed in 
section \ref{sec:modelcircuit} to interpret the time-dependent signals. In section 
\ref{sec:position-results} heating-position dependent measurements are presented, which 
reveal that the Seebeck voltage is generated locally at the MTJ in this geometry.

\section{Description of the experimental setup}
\label{sec:experimental-setup}
The setup used in this work is based on the experiments performed by Gravier et 
al.\cite{Gravier2004} on metallic nanowires. We adapted the electrical and optical 
techniques to the requirements for measuring small thermovoltages across a 
micrometer-sized single MTJ.

\begin{figure}[tb]
\includegraphics[width=0.95\linewidth]{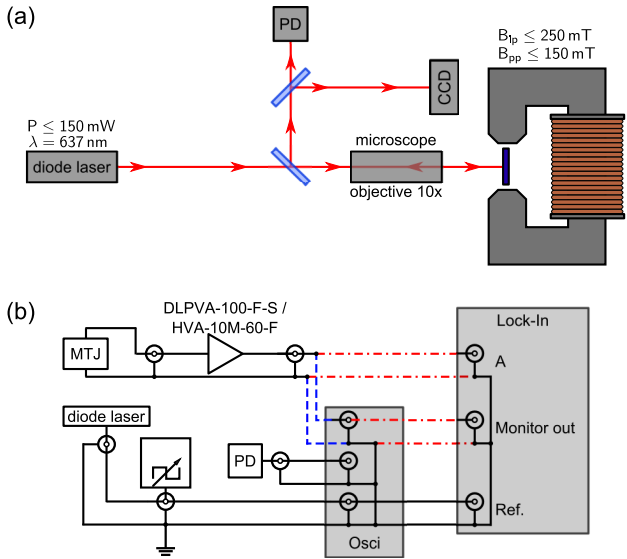}
\caption{Experimental setup for measurements of the TMS effect: (a) Optical setup with 
confocal microscope including a fast photodiode (PD) and a CCD camera, (b) electrical 
setup showing the connections of the sample and optoelectronic components to the lock-in 
amplifier and oscilloscope.}
\label{fig:exp-setup}
\end{figure}

To heat the MTJ from the top and to create a temperature gradient across the layer 
stack, a $150\,\mathrm{mW}$ laser diode (Toptica ibeam-smart-640-s) is focused down to a 
beamwaist of $5\,\mu\textrm{m}$ -- $10\,\mu\textrm{m}$ using a microscope objective 
(Mitutoyo M Plan Apo 10x). The central laser wavelength is 637~nm. An exact positioning 
of the laser spot onto the MTJ is crucial for obtaining reliable voltage measurements. 
Thus, the position of the laser spot can be controlled using a confocal microscope as 
depicted in FIG.~\ref{fig:exp-setup}~(a). With a set of different electromagnets, the 
sample can be studied in magnetic fields $B_{\mathrm{ip}} \leq 250\,\mathrm{mT}$ 
in-plane and $B_{\mathrm{pp}} \leq 150\,\mathrm{mT}$ perpendicular to plane.

The thermovoltage is detected with a lock-in amplifier. In our earlier 
publication\cite{Walter2011}, a mechanical chopper was used to modulate the laser 
heating at $1.5\,\mathrm{kHz}$, however it was found that the beamwaist of the 
unfocussed laser in combination with the mechanical chopping decreases the 
time-resolution of the thermovoltage detected by the oscilloscope. As a consequence, a 
waveform generator (Agilent 33500B) has been implemented to modulate the laser diode 
power with a square wave of $1.5\,\mathrm{kHz}$. As can be seen from 
FIG.~\ref{fig:exp-setup}~(a), a fast photodiode (EOT ET-2030, rise time of 
$<300\,\mathrm{ps}$) is integrated in the confocal microscope part of the setup to check 
the square-wave form of the light intensity. With this optical setup a rise and fall 
time of $< 1\,\mu\mathrm{s}$ of the light intensity can be achieved. This is faster than 
the rise time of the preamplifier used for high impedance MTJs. Thus, the 
time-resolution of the measured voltage signal of the MTJs is only limited by the 
electronic equipment,
which is shown in FIG.~\ref{fig:exp-setup}~(b): The MTJ is connected to a precision 
voltage preamplifier via shielded cables (total length 1.6~m). Depending on the MTJ 
resistance either a high impedance amplifier (femto DLPVA-100-F-S) with a rise time of 
$3.5\,\mu\mathrm{s}$ or a faster amplifier for low impedance sources (HVA-10M-60-F) 
with a rise time of $3.5\,\mathrm{ns}$ can be used. The signal preamplified by 
$60\,\mathrm{dB}$ to $80\,\mathrm{dB}$ is then fed either to the oscilloscope directly 
(blue dashed line in FIG.\ref{fig:exp-setup}~(b)) or to the lock-in amplifier (red 
dash-dotted line in FIG.\ref{fig:exp-setup}~(b)). In the latter case, the time-dependent 
voltage signals are recorded by the oscilloscope which then is connected to the monitor 
out of the lock-in amplifier. A Stanford Research Systems SR830 lock-in amplifier and a 
Philips PM3382 oscilloscope are used. All electrical components are triggered by the 
waveform generator and carefully grounded to minimize noise coupling into the 
measurement circuit. The noise level of the setup is within a range of $10\,\mathrm{nV}$ 
to $50\,\mathrm{nV}$, which is of the same order as Johnson-Nyquist-noise of the MTJ's 
resistance at room temperature. The low noise level enables observing small 
voltage changes on the order of nanovolts at low laser intensities, as shown in 
FIG.~\ref{fig:tmr-tms}.

For a thermocurrent measurement, the MTJ is connected without preamplifier to the 
lock-in amplifier set to current detection.

\subsection{Determination of temperatures}
\label{sec:temperature-determination}
Since the CoFeB and MgO layers of only a few nanometer thickness are buried under 
several layers of electrical leads, it is very difficult to reliably measure the 
temperature difference across them. Putting thermocouples to the electrical leads of the 
MTJ can only give a rough estimate of the temperature distribution.

\begin{table}
\caption{Material parameters for COMSOL simulations. If not specified otherwise, the 
values are taken from Refs.~\onlinecite{Walter2011,Liebing2012,Papusoi2008,Beecher1994}. 
The thermal conductivities used in the simulations are printed in bold letters. 
Experimental thin film values are given where available.}
\label{tab:comsol-material-parameters}
\begin{ruledtabular}
\begin{tabular}{c*{3}{c}}
Material & $\rho\,(10^3\frac{\text{kg}}{\text{m}^3})$ & 
$c_{V/p}\,(\frac{\text{J}}{\text{kg}\cdot\text{K}})$ & 
$\kappa_{\mathrm{bulk}}\,/\, 
\kappa_{\mathrm{thin}}^{\mathrm{exp}}\,(\frac{\text{W}}{\text{m}\cdot\text{K}})$ 
\\[0.5ex]\hline 
& & & \\[-2ex]
Au        & 19.32 & 128  & \textbf{320.0}\,/\, 70\footnote{Ref.~\onlinecite{Zink2005}} 
-- 170\footnote{Ref.~\onlinecite{Zhang2006}} \\
Cr        & 7.15  & 449  & \textbf{94.0}   \\
Ru        & 12.37 & 238  & \textbf{117.0}  \\
Ta        & 16.65 & 140  & \textbf{57.0}   \\
Permalloy & 8.7   & 460  & \textbf{19.0}   \\
IrMn      & 10.18 & 69.7 & \textbf{6.0}    \\
Co-Fe-B   & 8.22  & 440  & \textbf{86.7}   \\
MgO       & 3.58  & 935  & 48.0\,/\, \textbf{4.0}\footnote{Ref.~\onlinecite{Lee1995}} \\
SiO$_2$   & 2.20  & 1052 & \textbf{1.4}    \\
Si        & 2.33  & 700  & \textbf{150.0}  \\
SiN       & 3.11  & 700  & \textbf{35.9}   \\
\end{tabular}
%\label{tab:comsol-params}
%}
%\begin{tabnote}
%$^{\text a}$ The heat-conductivity of MgO is an experimentally found thin film 
%value\cite{Lee1995}.
%\end{tabnote}
\end{ruledtabular}
\end{table}
Consequently, the heat conduction equation is, therefore, numerically solved using 
COMSOL Multiphysics\cite{COMSOL} to determine the temperatures of the MTJ layers. The 
results have to be regarded as estimates, since interface heat resistances are not taken 
into account and bulk values of thermal conductivities, densities and heat capacities 
are used for the metals layers. These material parameters are taken from 
literature shown in TABLE~\ref{tab:comsol-material-parameters} for the MTJ materials. 
The table shows that the thermal conductivity of an Au thin film is lower by at least a 
factor of 2. In addition, we used the experimentally observed thin film value for the 
MgO layer, because here the thermal conductivity changes by an order of magnitude. This 
improves the reliability of our simulations.\cite{Lee1995,Walter2011} Further details 
on the simulations can be found in earlier 
publications.\cite{Walter2011,Leutenantsmeyer2013} The temperature difference across the 
$1.5\,\mathrm{nm}$ MgO layer resulting from the simulations is used in 
section~\ref{sec:TMR-TMS-results} in combination with the measured voltage to calculate 
the Seebeck coefficients.

\section{Sample preparation}
\label{sec:sample-preparation}
The MTJs are prepared on two types of substrates: MgO and oxidized p-type silicon (Si) 
($50\,\mathrm{nm}$ $\mathrm{SiO_2}$, resistivity of $20\,\mathrm{\Omega cm}$) by sputter 
deposition in a Leybold Vakuum GmbH CLAB 600. The film system on MgO consists of bottom 
contact Ta 5/Ru 30/Ta 10/Ru 5; pinned layer MnIr 15/CoFeB 3; tunnel barrier MgO 1.5; 
free layer CoFeB 3/NiFe 6; top contact Ta 3/Ru 3/Ta 3/Au 15 (thickness are given in nm). 
In case of Si/$\mathrm{SiO_2}$ substrates the pinned layer is slightly changed to MnIr 
12/CoFe 3/Ru 0.9/CoFeB 3. Elliptical MTJs with a size of $6\,\mathrm{\mu m}\times 
4\,\mathrm{\mu m}$ are produced by electron beam lithography and subsequent ion beam 
etching. Afterwards, 100~nm of SiN are sputter deposited next to the MTJs as insulator. 
An Au bond pad is placed adjacent to the MTJs in an additional sputtering and patterning 
process for connecting the 15 nm Au top contact to the measurement electronics. This 
allows free optical access to the MTJ.

\section{Experiments on Magnesium oxide and Silicon substrates}
\label{sec:experiments}

\subsection{TMR and TMS measurements}
\label{sec:TMR-TMS-results}
Thermoelectric effects can be derived theoretically from thermodynamic principles. For 
the case of an MTJ, the thermoelectric coefficients dependent on the tunneling 
probability can be expressed by equations based on the Landauer 
formula.\cite{Callen1960,Buettiker1985,Sivan1986,Czerner2011,Lin2012} In this way, the 
influence of spin transport on the Seebeck voltage of an MTJ can be described. To 
clarify the interconnection between the different transport coefficients and sign 
conventions, we first derive the Seebeck voltage, Seebeck current, TMS and TMR from the 
thermodynamic kinetic equations and the moments of the transport integral.

The charge transport through the barrier of the MTJ is given as
\begin{equation}
I = G V + G S \Delta T \label{eq:transport}
\end{equation}
where $G$ is the electric conductance and $S$ is the Seebeck coefficient. According to 
eq. \ref{eq:transport}, a current $I$ is either generated by an external voltage $V$ or 
by a temperature gradient $\Delta T$. In a Seebeck current measurement no external 
voltage is applied to the MTJ ($V=0$) whereas in a perfect voltage measurement no 
current is transported in the circuit ($I=0$), which yields 
\begin{equation}
	I = G S \Delta T, \qquad V = -S \Delta T 
	\label{eq:current_voltage}
\end{equation}
for the measured current and voltage, respectively. The coefficients can be rewritten 
as\cite{Ouyang2009,Czerner2011}
\begin{equation}
	G = e^2 L_0, \qquad S = -\frac{1}{e T}\frac{L_1}{L_0} \label{eq:conductance}
\end{equation}
using the moments
\begin{equation}
L_n=\frac{2}{h} \int T\left(E\right)\left(E-\mu \right)^n \left[-\partial_E 
f\left(E,\mu,T\right)\right]\mathrm{d}E
\label{eq:moments}
\end{equation}
dependent on $f\left(E,\mu,T\right)$, the Fermi occupation function at a given energy 
$E$, electrochemical potential $\mu$ and temperature $T$ and on the energy-dependent 
transmission probability $T\left(E\right)$ that is different for the parallel (P) and 
antiparallel (AP) orientation of the bottom and top layer's magnetization, which leads 
to different moments for both states\cite{Czerner2011,Walter2011}. Thus, the TMS is 
calculated analogous to the TMR:
\begin{equation}
\mathrm{TMR} = 
\frac{R_\mathrm{AP}-R_\mathrm{P}}{R_\mathrm{P}}, 
\qquad
\mathrm{TMS} = 
\frac{S_\mathrm{P}-S_\mathrm{AP}}{\min\left(|S_\mathrm{P}|,|S_\mathrm{AP}|\right)}.
\end{equation}

\begin{figure}[t]
\includegraphics[width=0.95\linewidth]{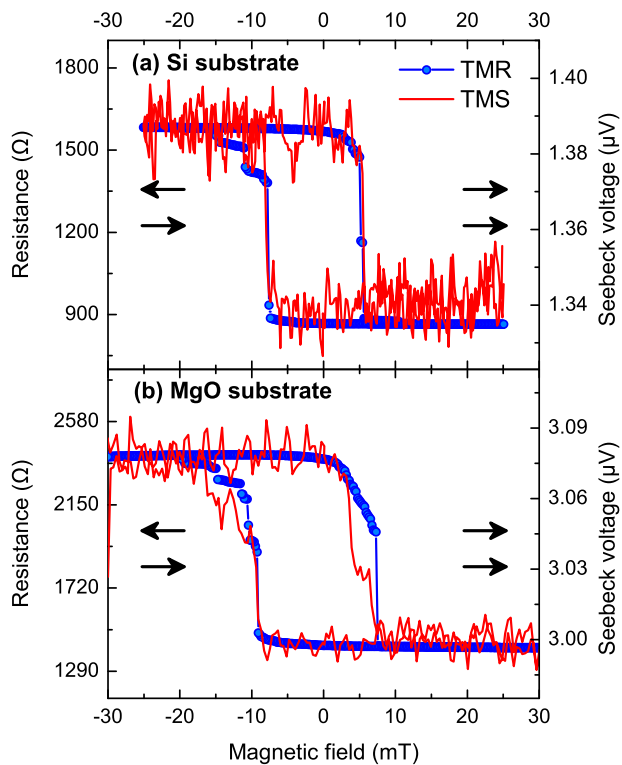}
\caption{Low laser power and different substrates: TMR (blue circles, left 
scale) and TMS (red line, right scale) of nominally identical MTJs on (a) 
Si/$\mathrm{SiO_2}$ and (b) MgO substrates obtained with a laser power of 10~mW and 
15~mW, respectively. The arrows represent the relative orientation of the magnetic 
layers. 
The TMR and TMS values are given in TABLE \ref{tab:tmr-tms}.}
\label{fig:tmr-tms}
\end{figure}

The MTJs on Si/$\mathrm{SiO_2}$ and MgO were prepared to investigate the influence of 
the substrate material on the TMR and TMS measurements. FIG.~\ref{fig:tmr-tms} shows 
field dependent resistance and Seebeck voltage curves of elliptical MTJs with an area of 
19 $\mathrm{\mu m}^2$ prepared on Si/$\mathrm{SiO_2}$ and MgO, respectively.

In case of the Si/$\mathrm{SiO_2}$ substrate, the resistance of the MTJ switches between 
$1583\,\mathrm{\Omega}$ in the antiparallel and $864\,\mathrm{\Omega}$ in the parallel 
orientation of the ferromagnetic layers, which yields a TMR ratio of $83\,\%$. The 
Seebeck voltage, generated by laser heating with a power of 10~mW, changes from 
$1.39\,\mathrm{\mu V}$ in the antiparallel to $1.34\,\mathrm{\mu V}$ in the parallel 
state resulting in a TMS ratio of $3.7\,\%$. The Seebeck voltage detected by the lock-in 
amplifier is positive as shown in FIG.~\ref{fig:Si-MgO-osci}~(a). This means that the 
electrons are accumulated at the cold electrode, which results in a negative Seebeck 
coefficient (eq.~(\ref{eq:current_voltage})).

The MTJ on MgO exhibits a larger resistance than the MTJ on Si/$\mathrm{SiO_2}$ 
substrate. The resistance varies between $2400\,\mathrm{\Omega}$ in the antiparallel 
and $1411\,\mathrm{\Omega}$ in the parallel state. A TMR ratio of $70\,\%$ is obtained. 
The corresponding Seebeck voltage, induced by laser heating with 15~mW laser power, 
switches at the same magnetic fields between $3.08\,\mathrm{\mu V}$ and $3.00\,
\mathrm{\mu V}$ gaining a TMS ratio of $2.6\,\%$. As for the MTJ on Si/$\mathrm{SiO_2}$ 
substrate, the Seebeck voltage is again positive (FIG.~\ref{fig:Si-MgO-osci}~(b)).

In TABLE \ref{tab:tmr-tms} the parameters of the TMR and TMS are shown. The small 
differences are within the normal deviations between different MTJs and can also be due 
to different growth conditions on the two substrates. As a consequence, no evidence for 
influence of parasitic Seebeck voltages arising from different substrates on the TMS 
measurements is found.

The Seebeck coefficients in TABLE \ref{tab:tmr-tms} are calculated from the above 
mentioned simulated temperature gradient across the barrier and the thermovoltage 
generated inside the MTJ. This thermovoltage consists of a spin-dependent component from 
the ferromagnetic electrodes and a spin-independent background from the other layers in 
the MTJ. A possible solution to estimate this spin-independent background is given in 
Ref.\onlinecite{Liebing2012}: The MTJ is forced to a dielectric breakdown after the TMS 
measurement is performed and the remaining, spin-independent background thermovoltage is 
determined. The background thermovoltage is approximately 
$0.05\,\frac{\mu\mathrm{V}}{\mathrm{mW}}$ up to 
$0.4\,\frac{\mu\mathrm{V}}{\mathrm{mW}}$, such that after subtraction, the resulting TMS 
ratios are around 20 \% for the data presented in this paper. However, the morphology 
of the layers changes due to the voltage stress applied to the MTJ, e.g. the CoFeB can 
change from an amorphous to a crystalline structure and the interfaces between 
the thin films can be destroyed\cite{Appl.Phys.Lett.-93-152508}. Therefore, this 
method allows only an estimation for the background thermovoltages arising from other 
sources of the layer stack of the tunnel junction that do not contribute to the TMS 
itself.
\begin{table}[h]
\caption{Comparison of TMR and TMS on MgO and Si/$\mathrm{SiO_2}$ samples}
\label{tab:tmr-tms}
\begin{ruledtabular}
\begin{tabular}{c c c c}
 substrate & $R_\mathrm{P} \left(\mathrm{\Omega}\right)$ & $R_\mathrm{AP} 
 \left(\mathrm{\Omega}\right)$ & TMR \\[0.5ex]\hline
Si  & 864  & 1583 & 83 \% \\[0.5ex] %K120531a-F07c-15mW-Si
MgO & 1411 & 2400 & 70 \% \\ %K120702a-10d-10mW-MgO
\end{tabular}
\begin{tabular}{c c c c c c}
substrate & $V_\mathrm{P} \left(\mathrm{\mu V}\right)$ & $V_\mathrm{AP} 
\left(\mathrm{\mu V}\right)$ & $S_\mathrm{P} \left(\frac{\mu 
\mathrm{V}}{\mathrm{K}}\right)$\footnote{$\Delta T_\mathrm{MgO} = 6\,\mathrm{m 
K}$ is used for Si/$\mathrm{SiO_2}$ substrate and $\Delta T_\mathrm{MgO} = 4\,\mathrm{m 
K}$ for MgO substrate.} & $S_\mathrm{AP} \left(\frac{\mu 
\mathrm{V}}{\mathrm{K}}\right)$\footnotemark[1] & TMS \\[0.5ex]\hline
Si  & 1.34 & 1.39 & -223 & -232 & 3.7 \% \\[0.5ex] %K120531a-F07c-15mW-Si
MgO & 3.00 & 3.08 & -750 & -770 & 2.6 \% \\ %K120702a-10d-10mW-MgO
\end{tabular}
\end{ruledtabular}
\end{table}

\subsection{Thermocurrent measurements}
\label{sec:thermocurrent-results}
\begin{figure}[th]
	\includegraphics[width=0.91\linewidth]{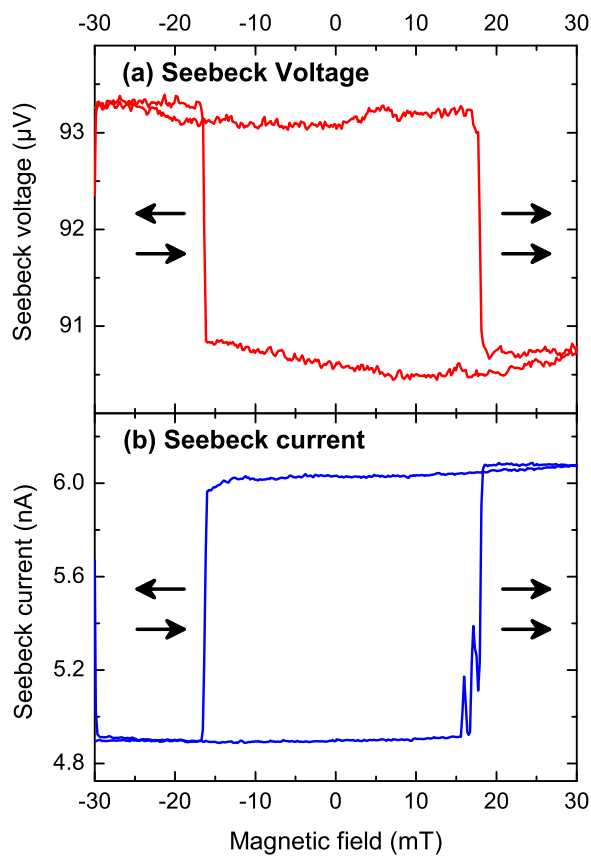}
	\caption{High laser power: Seebeck voltage (a) and Seebeck current (b) 
	measured at a laser power of 150 mW.}
	\label{fig:current}
\end{figure}

In an open circuit, the Seebeck effect creates a voltage in an MTJ experiencing a 
temperature gradient, whereas in a closed circuit geometry, it can drive a Seebeck 
current. FIG.~\ref{fig:current} shows the magnetization dependence of the Seebeck 
voltage and Seebeck current induced by laser heating with a power of 150 mW for an MTJ 
on MgO with an area of $1.57\,\mathrm{\mu m}^2$ and a resistance of $28.1\,\mathrm{k 
\Omega}$ in the antiparallel and $16.7\,\mathrm{k \Omega}$ in the parallel state. Note 
that the laser power is considerably larger than in the first example 
(FIG.~\ref{fig:tmr-tms}). The voltage varies between $93.30\,\mathrm{\mu V}$ in the 
antiparallel and $90.72\,\mathrm{\mu V}$ in the parallel state, whereas the current 
behaves inversely such that it reaches $4.90\,\mathrm{nA}$ and $6.07\,\mathrm{nA}$, 
respectively. This yields a TMS ratio of $2.84\,\%$ and a current effect-ratio of 
$23.9\,\%$.

Since the moments in eq.~(\ref{eq:moments}), which depend on the magnetization dependent 
transmission $T(E)$, occur in the conductance $G$ as well as in the Seebeck coefficient 
$S$ (eq.~(\ref{eq:conductance})), both, the voltage and the current should exhibit a 
magnetic field dependent variation, as suggested by eq.~(\ref{eq:current_voltage}). This 
prediction is confirmed by our experimental results. The difference in the effect 
amplitudes is explained by the fact that, as it can be seen in 
eq.~(\ref{eq:current_voltage}), the voltage only depends on the Seebeck coefficient $S$, 
whereas the current is additionally dependent on the electrical conductance $G$, which 
is strongly dependent on the magnetization alignment due to the high TMR ratio.

Seebeck currents were also investigated by Lin et al.\cite{Lin2012}, but in contrast to 
our measurements they do not detect a dependency of the current on the magnetic field 
which they explain by the different mechanisms causing TMS in alumina-based MTJs. 
On Co-Fe-B/MgO MTJs, Liebing et al. demonstrated magnetic switching in Seebeck 
current measurements very recently.\cite{Liebing2013}

\subsection{Time-dependent thermovoltage signals}
\label{sec:osci-results}
To gain a deeper understanding of the processes leading to the TMS signal measured by 
the lock-in amplifier, a closer investigation of the time-dependent voltage signal is 
essential. It is assumed that the temperature gradient rapidly increases and decreases 
upon laser on/off, which is justified by temperature simulations yielding a time of $< 
2\,\mu\mathrm{s}$ to reach equilibrium. Thus, a nearly rectangular time-dependent 
voltage signal is expected corresponding to the laser modulation.

\begin{figure}[t]
\includegraphics[width=0.95\linewidth]{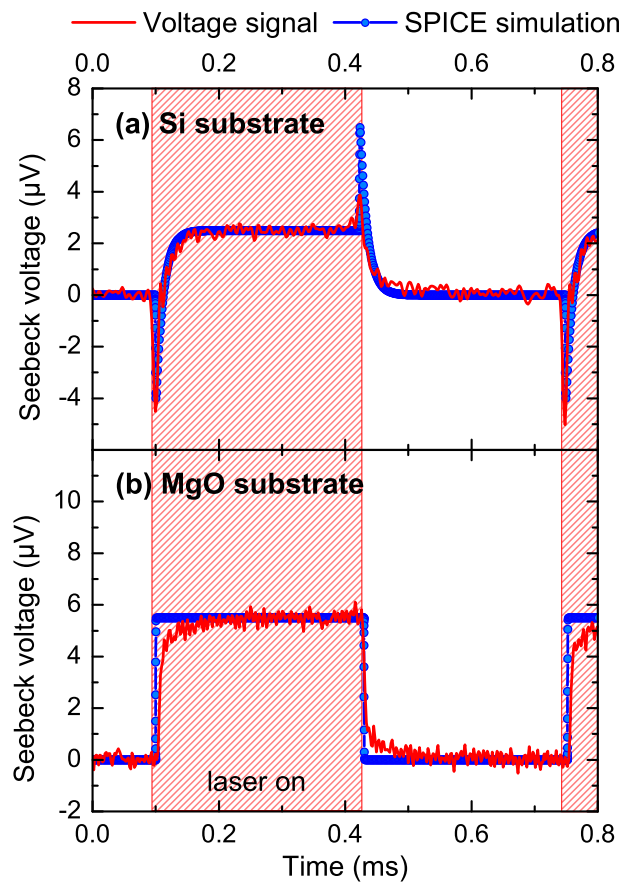}
\caption{Time-dependent voltage signals of MTJs on (a) Si/$\mathrm{SiO_2}$ and 
(b) MgO substrate with a laser power of 10 mW and 15 mW, respectively. As blue circles 
Simulation Program with Integrated Circuit Emphasis (SPICE) simulations are shown, as 
described in section \ref{sec:modelcircuit}.}
\label{fig:Si-MgO-osci}
\end{figure}

In FIG.~\ref{fig:Si-MgO-osci}, the time-dependent voltage signals of MTJs on 
Si/$\mathrm{SiO_2}$ and MgO are depicted. The traces for both substrate materials 
clearly reveal voltage plateaus with a small rise and fall-time when the laser is turned 
on and off. Whereas on MgO the rectangular shape is clearly visible, unexpected 
negative and positive voltage peaks can be additionally observed at the start and end of 
the heating period on Si/$\mathrm{SiO_2}$.
The position and shape of these voltage peaks suggest an electrical capacitance 
as their origin. The source can be further restricted to the substrate as the 
additional voltage only occurs in samples with p-doped Si substrate which is 
capacitively coupled to the bottom electrode by the $50\,\mathrm{nm}$ $\mathrm{SiO_2}$ 
dielectric.

\section{Development of a model circuit}
\label{sec:modelcircuit}
\begin{figure}[ht]
\centering
\includegraphics[width=0.95\linewidth]{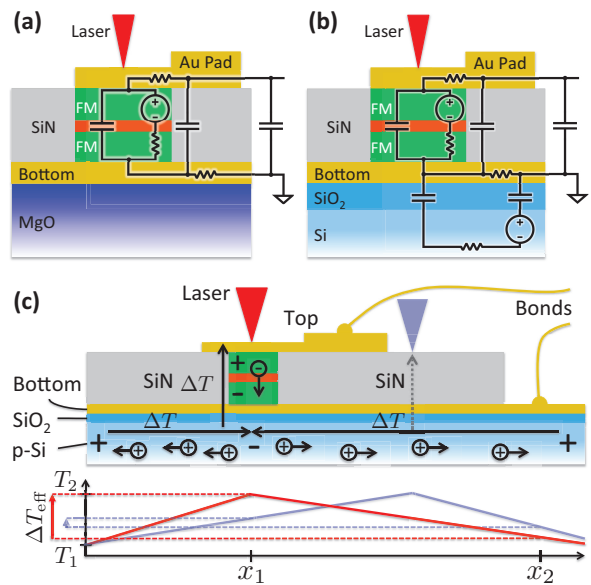}
\caption{Model circuit for MTJs on (a) insulating MgO and (b) capacitively coupled 
p-type Si substrates. (c) Inside the samples on Si two heat gradients $\Delta 
T_{\mathrm{MTJ}}$ and $\Delta T_{\mathrm{Si}}$ produce thermovoltages 
$V_{\mathrm{MTJ}}$ and $V_{\mathrm{Si}}$, respectively. When the laser is positioned on 
the MTJ (red) the effective temperature gradient $\Delta T_\mathrm{eff}$ between the 
contact points $x_1$ and $x_2$ is larger compared to the laser positioned between the 
MTJ and the edge of the sample (blue).}
\label{fig:model}
\end{figure}

Uncovering the processes responsible for the strikingly different temporal voltage 
traces measured on MgO and Si/$\mathrm{SiO_2}$ substrates can be achieved by describing 
the sample structure as a model circuit. FIG.~\ref{fig:model} sketches how the relevant 
parts inside the sample can be converted into an equivalent circuit consisting of three 
major units.
\begin{table}[ht]
	\centering
	\caption{Basic estimations for calculating the resistance and capacitance in the 
	model circuit for simulations.}
	\label{tab:calculations}
	\begin{ruledtabular}
	\begin{tabular}{p{0.15\linewidth} p{0.6\linewidth} r}
	\textbf{comp.} & \textbf{how to calculate} & \textbf{value}\\[0.5ex]\hline
	 & \\[-1ex]
	$R_\mathrm{MTJ}$    & extracted from TMR data & \\[0.5ex]
	$V_\mathrm{MTJ}$    & fit to data (value of plateau)& \\[0.5ex]
	$V_\mathrm{Si}$     & fit to data & \\[0.5ex]
	$C_\mathrm{MTJ}$    & parallel plate capacitor: MgO thickness, MTJ area & 
	$1.08\,\mathrm{pF}$\\[0.5ex]
	$R_\text{Au-pad}$   & geometric dimensions of Au pad and resistivity of Au & 
	$10\,\Omega$\\[0.5ex]
	$C_\text{Au-pad}$   & parallel plate capacitor: SiN thickness, area of bond pad & 
	$2.8\,\mathrm{pF}$\\[0.5ex]
	$C_\text{cables}$   & given by manufacturer & $160\,\mathrm{pF}$\\[0.5ex]
	$C_\mathrm{SiO_2}$  & effective parallel plate capacitor with dielectric 
	$\mathrm{SiO_2}$ & $70\,\mathrm{nF}$\\[0.5ex]
	$R_\mathrm{Si}$     & geometric dimension of Si substrate, conducting channel is 
	created between MTJ and bond wire contact to bottom electrode & 
	$300\,\Omega$\\[0.5ex]
	$R_\mathrm{bottom}$ & geometric dimensions of conducting channel inside the bottom 
	electrode between MTJ an bond wire contact & $40\,\Omega$\\
	\end{tabular}
	\end{ruledtabular}
\end{table}

The first part is the MTJ simplified as a voltage source $V_\mathrm{MTJ}$ simulating the 
Seebeck voltage generated by the temperature gradient across the barrier, a resistor 
$R_\mathrm{MTJ}$ representing the dielectric barrier and a capacitor $C_\mathrm{MTJ}$ 
describing the capacitance built up by the two ferromagnetic layers (FM) separated by 
the MgO.

The second unit contains the electrodes and wiring including the resistance of the 
bottom electrode $R_\mathrm{bottom}$ and of the top contact, mainly the gold bond pad 
$R_\text{Au-pad}$. Furthermore a capacitance $C_\text{Au-pad}$ is build up by the gold 
pad and the bottom electrode separated by the insulator $\mathrm{SiN}$ surrounding the 
MTJs. It is supplemented by the cable capacitance $C_\mathrm{cable}$ of the coaxial 
cables connecting the sample to the electronic equipment.

The third major component is the substrate. In case of MgO samples, the substrate is 
insulating and therefore has not to be taken into account when constructing a model 
circuit (FIG.~\ref{fig:model}~(a)). When Si samples are depicted, the substrate is a 
p-type semiconductor that creates a Seebeck voltage $V_\mathrm{Si}$ when heated (FIG.~
\ref{fig:model}~(b)) that capacitively couples to the bottom electrode through the 
$\mathrm{SiO_2}$ capping. The temperature gradient inside the substrate arises 
when the MTJ is heated. Not only the upper side of the MTJ is heated when irradiated by 
the laser, but also the lower part near the substrate experiences an elevated 
temperature due to heat conduction through the layer stack. Thus, the temperature inside 
the substrate underneath the MTJ is higher than at the edges of the sample. The 
resulting temperature gradient creates a Seebeck voltage in the p-Si substrate as 
sketched in FIG.~\ref{fig:model}~(c). This effect is included in the model circuit 
by adding a voltage source and two capacitances $C_\mathrm{SiO_2}$, one underneath the 
MTJ and another at the point where the bottom electrode is connected to the gold bond 
wire, as these two points confine the segment where a capacitively coupled voltage can 
be detected (FIG.~\ref{fig:model}~(b)).

The size of all capacitances can be calculated from the model of a parallel plate 
capacitor except the cable capacitance that is given by the manufacturer. The sum of the 
resistors $R_\mathrm{MTJ}$, $R_\mathrm{bottom}$ and $R_\text{Au-pad}$ connected in 
series has to match the measured TMR whilst the value of $R_\mathrm{bottom}$ and 
$R_\mathrm{Si}$ can be deduced from the geometric dimensions of the conduction channel 
constituting between the MTJ and the contact point of the bottom electrode to the gold 
bond wire.

FIG.~\ref{fig:Si-MgO-osci} shows a good agreement of SPICE simulations of the model 
circuit with the measured data. Our model of a Seebeck voltage created in the silicon 
substrate that cannot occur inside the MgO substrate explains the effects observed in 
the experiment. The absolute values of the voltages $V_\mathrm{MTJ}$ and $V_\mathrm{Si}$ 
have to be deduced. All other components are calculated and summarized in 
TABLE~\ref{tab:calculations}. Considering that the relation of the amplitude $A$ 
detected by the oscilloscope and the voltage output $V$ of the lock-in amplifier is 
given by $V \approx 0.5\cdot A$ for a square wave signal, the measurements of 
FIG.~\ref{fig:Si-MgO-osci} agree very well with those of FIG.~\ref{fig:tmr-tms}. Based 
on these results, we are able to assign the different measured voltages to a Seebeck 
effect inside the MTJ on the one hand and a Seebeck effect inside the substrate material 
on the other hand.

\section{Position-dependent measurements}
\label{sec:position-results}
The tunnel magneto-Seebeck effect should arise from Seebeck voltages generated by a 
temperature gradient across the MgO layer sandwiched between two ferromagnets. From the 
underlying geometry of the MTJ sketched in FIG.~\ref{fig:model}~(c) we conclude that 
moving the laser spot away from the junction on the gold bond pad should already 
decrease this temperature gradient and hence the observed Seebeck voltage. Furthermore, 
lateral temperature gradients created by the laser spot on the gold bond pad should 
cancel out because of the lateral radial symmetry of the heating.

\begin{figure*}[ht]
\includegraphics[width=0.9\linewidth]{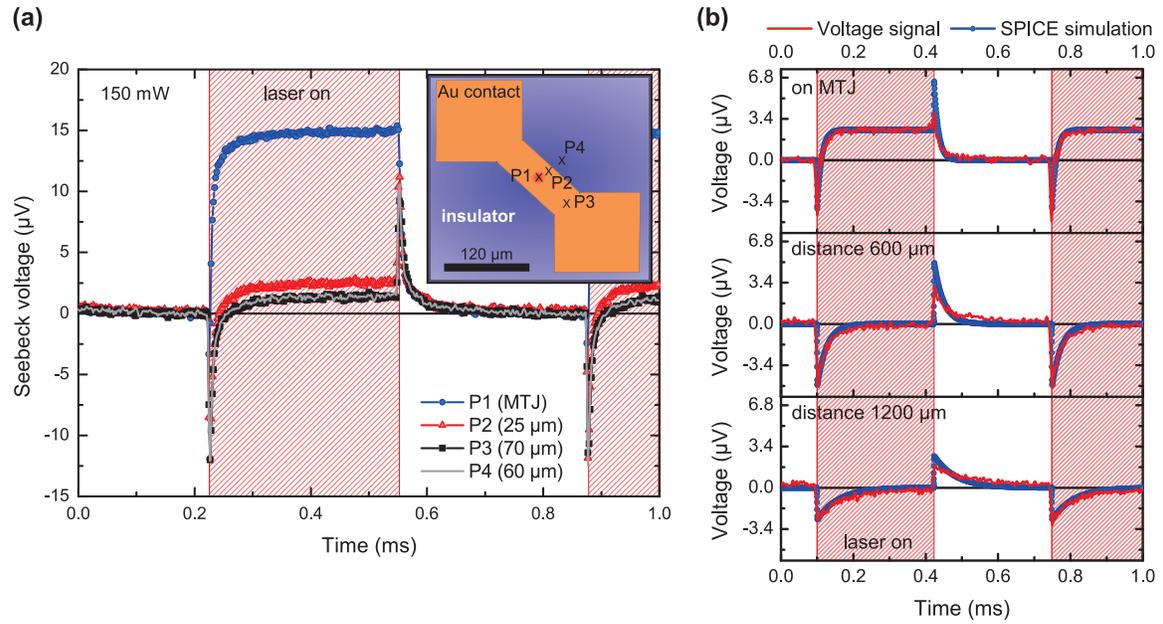}
\caption{(a) Short distances: Measurements in the vicinity of the MTJ; the 
positions are shown in the inset. Laser spot and MTJ are located at P1, the heated area 
is sketched in red. (b) Large distances: Measurements of a different MTJ at distances 
more than factor of 10 larger as in the case of (a).}
\label{fig:position-dependence}
\end{figure*}
We performed several position-dependent measurements in the vicinity of the gold bond 
pad of the junction, and also with the laser spot moved more than $1\,\mathrm{mm}$ away 
from the MTJ. The inset of FIG.~\ref{fig:position-dependence}~(a) shows the geometry of 
the bond pad with the MTJ located at position P1. The area heated by the laser is 
sketched as a red circle. Its diameter is according to simulations\cite{COMSOL} only 
slightly larger than the laser spot ($d=10\,\mu\mathrm{m}$). TMS measurements were taken 
with the laser positioned at P1 -- P4. The corresponding time-dependent voltage signals 
are shown in FIG.~\ref{fig:position-dependence}~(a). It can be seen that a 
square-wave-like Seebeck voltage that is attributed to the MTJ, as discussed in 
Secs.~\ref{sec:experiments}--\ref{sec:modelcircuit}, occurs only if the laser spot is 
centered directly onto the MTJ at P1. Only at 
this position the TMS effect is observed. At position P2, adjacent to the MTJ, the 
Seebeck voltage of the MTJ is already strongly reduced. Only the voltage peaks 
attributed to voltages generated in the Si substrate occur at all four positions. 
Time-dependent voltage signals were recorded also for large distances, which are shown 
in FIG.~\ref{fig:position-dependence}~(b): It is observed that the time constant of the 
exponential decay of the voltage peaks increases with distance. The plotted simulated 
curves describe the signals reasonably well. In the model, the MTJ voltage 
$V_\mathrm{MTJ}$ is set to zero for non-zero distances between the laser and the MTJ and 
the voltage peaks can be reproduced by adjusting the resistance of the substrate and by 
lowering the generated voltage (reduced effective temperature gradient, 
FIG.~\ref{fig:model}~(c)) according to the increased distance.

These findings support the attribution of the voltage peaks to parasitic voltages of the 
Si/$\mathrm{SiO_2}$ substrate. Further, they show that the setup enables us to 
discriminate voltages locally generated in a single MTJ.

\section{Conclusion}
In summary, we have presented an experimental setup that allows the reliable measurement 
of thermomagnetoelectric effects in a single tunnel junction with nanovolt resolution at 
a temporal resolution of a few microseconds. MTJs grown on oxidized Si and MgO 
substrates have been tested and show comparable Seebeck voltages and currents. We find a 
magnetic effect also in the Seebeck current measurements. Further, with the improved 
temporal resolution, the voltage signals of the MTJs can be interpreted with the help of 
a model circuit. On oxidized Si substrates an additional voltage generated in the 
substrate can be identified. However, no evidence is found that this voltage influences 
the TMS measurements carried out with a lock-in amplifier. Distance dependent 
measurements reveal that the detected Seebeck voltage originates only from the MTJ layer 
stack. When the laser is moved away from the MTJ, only the Seebeck voltage signal of the 
substrate can be found.

\begin{acknowledgments}
M.M., A.T., and G.R. are supported by the DFG through SPP~1538 SpinCaT (MU1780/8-1, 
RE1052/24-1). V.D. and A.T. are supported by the Ministry of Innovation, Science and 
Research (MIWF) of North Rhine-Westphalia with an independent researcher grant.
\end{acknowledgments}

%\appendix
%\section{Appendixes}

%%\nocite{*}
\bibliography{Boehnke-Manuscript.bib}% Produces the bibliography via BibTeX.

\end{document}